\documentclass[fleqn,12pt,twoside]{article}
\usepackage{espcrc1}

\usepackage{graphicx}
\usepackage[figuresright]{rotating}

\newcommand{\AmS}{{\protect\the\textfont2
  A\kern-.1667em\lower.5ex\hbox{M}\kern-.125emS}}
\newcommand{\simge}{\hspace*{0.2em}\raisebox{0.5ex}{$>$}
     \hspace{-0.8em}\raisebox{-0.3em}{$\sim$}\hspace*{0.2em}}

\newcommand{\beq}{\begin{equation}}
\newcommand{\eeq}{\end{equation}}
\newcommand{\bqa}{\begin{eqnarray}}
\newcommand{\eqa}{\end{eqnarray}}

\title{Few-Body Effects in Cold Atoms and Limit Cycles}

\author{H.-W. Hammer\address[HISKP]{Helmholtz-Institut f\"ur Strahlen-
und Kernphysik (Theorie), Universit\"at Bonn, 
Nussallee 14-16, 53115 Bonn, Germany}}%
       
\begin{document}

\maketitle

\begin{abstract}
Physical systems with a large scattering length have universal
properties independent of the details of the interaction at short
distances. 
Such systems can be realized in experiments with cold atoms close to
a Feshbach resonance. They also occur in many other areas of physics such
as nuclear and particle physics. The universal properties include a
geometric spectrum of three-body bound states (so-called Efimov states)
and log-periodic dependence of low-energy observables on the physical
parameters of the system. This behavior is characteristic
of a renormalization group limit cycle.
We discuss universality in the three- and four-body sectors and give an
overview of applications in cold atoms.
\end{abstract}

\section{INTRODUCTION}
The Effective Field Theory (EFT) approach provides a powerful framework 
that exploits the separation of scales in physical systems.
Only low-energy (or long-range) degrees of freedom are included
explicitly, with the rest parametrized in terms of the most general
local (contact) interactions. This procedure exploits the fact that
a low-energy probe of momentum $k$ cannot resolve structures on 
scales smaller than $R\sim 1/k$.\footnote{Note that $\hbar=1$ in this talk.}
Using renormalization,
the influence of short-distance physics on low-energy observables
is captured in a small number of low-energy constants.
Thus, the EFT describes universal low-energy physics independent of
detailed assumptions about the short-distance dynamics. 
All physical observables can be described in a controlled expansion in 
powers of $kl$, where $l$ is the characteristic low-energy 
length scale of the system. The size of $l$ depends on the system 
under consideration: for a finite-range potential, e.g., it is given 
by the range of the potential.
For the systems discussed here, $l$ is of the order of the effective 
range $r_e$ or the van der Waals scale $l_{vdW}$.

Effective Field Theories can be obtained by applying a renormalization
group (RG) transformation to a (more) fundamental theory. The RG 
transformation integrates out high momentum modes from the fundamental 
theory and leads to a description of low-energy physics in terms of low-energy
degrees of freedom only. It can be understood as a change of the resolution 
scale of the EFT. The transformation 
generates a flow in the space of coupling constants
that determine the effective Lagrangian ${\cal L}$. If the RG
transformation is continuous, this flow 
can be expressed by a differential equation for 
the coupling constants $g$:
\begin{equation}
\Lambda \frac{d}{d\Lambda} g = \beta(g)\,,
\label{eq:RGeq}
\end{equation}
where $\Lambda$ is the ultraviolet cutoff which corresponds to the
inverse resolution scale.
The function $\beta(g)$ determines the behavior of $g$ under
the RG transformation. The RG equation (\ref{eq:RGeq}) can have many types 
of solutions. The two simplest ones are (a) renormalization group
fixed points and (b) renormalization group limit cycles.
Fixed points are ubiquitous in condensed matter and particle physics.
They play an important role, e.g.,  in critical phenomena and 
the scale dependence of coupling constants in high energy physics.
Renormalization group limit cycles have been suggested by Ken Wilson
already in 1971 \cite{Wilson:1970ag}, 
but their phenomenological importance has only 
been realized very recently.

An important signature of a limit cycle is the appearance of a discrete
scaling symmetry, the invariance of physical observables under 
changes of the resolution scale by a discrete scale transformation 
with a preferred scaling factor $\lambda$. 
The prime example of a RG limit cycle is the Efimov effect \cite{Efi71}
for 3-body systems with large S-wave scattering length $a$.
Vitaly Efimov discovered that in the limit $a = \pm \infty$ there can be
infinitely many 3-body bound states (trimers) with an accumulation point 
at the scattering threshold \cite{Efi71}.  These trimers are called
Efimov states.  One remarkable feature of Efimov states is that
they have a geometric spectrum with preferred scaling factor 
$\lambda^2\approx 515$:
\begin{eqnarray}
E^{(n)}_T = \lambda^{2(n_*-n)} \hbar^2 \kappa^2_* /m,
\label{kappa-star}
\end{eqnarray}
where $\kappa_*$ is the binding wavenumber of the branch of Efimov states
labeled by $n_*$.  
The Efimov effect is just one example of the universal
phenomena characterized by the discrete scaling symmetry in the
3-body system \cite{Efi71}.
These universal properties persist also for finite
values of the scattering length as long as $a \gg l$.
For a review of these phenomena,
which we refer to as Efimov physics, see Ref.~\cite{Braaten:2004rn}.

For a generic system, the scattering length
is of the same order of magnitude as the low-energy length scale $l$.
Only a very specific choice of the parameters in the underlying theory 
(a so-called fine tuning) will generate a large scattering length.
Nevertheless, systems with large scattering length can be found in many
areas of physics. Examples are the S-wave scattering of nucleons
and of $^4$He atoms. For alkali atoms close to a Feshbach resonance,
$a$ can be tuned experimentally by adjusting an external magnetic field.
This is particularly interesting, since it allows to experimentally 
test the dependence of physical observables
on the scattering length.

In this talk, we discuss some applications of an EFT 
for few-body systems with large scattering length $a \gg l$
\cite{BhvK99}. In this theory, the limit cycle is manifest in the
RG behavior of the 3-body interaction which makes it ideally suited to 
study the Efimov effect and its consequences for other 3-body observables.

\section{THREE-BODY SYSTEM WITH LARGE SCATTERING LENGTH}

We first give a very brief review of the EFT
for few-body systems with large scattering length $a$, 
focusing on S-waves. A more detailed treatment is given in 
Ref.~\cite{Braaten:2004rn}.

For typical momenta $k\sim 1/a$,
the EFT expansion is in powers of $l/a$ so that higher
order corrections  are suppressed by powers of $l/a$. 
We consider a 2-body system of nonrelativistic bosons (referred to as atoms)
with large scattering length $a$ and mass $m$.
At sufficiently low energies, the most general Lagrangian 
may be written as:
\begin{equation}
{\cal L}  =  \psi^\dagger
             \left(i\partial_{t}+\frac{\vec{\nabla}^{2}}{2m}\right)\psi
 - \frac{C_0}{2} (\psi^\dagger \psi)^2
 - \frac{D_0}{6} (\psi^\dagger\psi)^3 + \ldots\,,
\label{eq:eftlag}
\end{equation}
where the $C_0$ and $D_0$ are nonderivative 2- and 3-body interaction
terms, respectively. The strength of the $C_0$ term is determined by
the scattering length $a$, while $D_0$ depends on a 3-body parameter
to be introduced below. The dots represent higher-order terms
which are suppressed at low-energies. For momenta $k$ of the order of
the inverse scattering length $1/a$, the problem is nonperturbative 
in $ka$. The exact 2-body scattering amplitude can be obtained
analytically by summing the so-called bubble diagrams with the
$C_0$ interaction term. The $D_0$ term does not contribute to 2-body
observables. After renormalization, 
the resulting amplitude reproduces the leading order of the 
well-known effective range expansion for the atom-atom
scattering amplitude:~$f_{AA}(k)=(-1/a -ik)^{-1}\,,$
where the total energy is $E=k^2/m$. 
If $a>0$, $f_{AA}$ has a pole at $k=i/a$ corresponding
to a shallow dimer with the universal binding energy $B_2=1/(ma^2)$. 
Higher-order interactions are perturbative and give the 
momentum-dependent terms in the effective range expansion.

\begin{figure}[htb]
\centerline{\includegraphics*[width=14cm,angle=0]{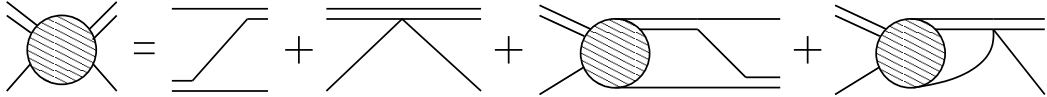}}
\caption{Integral equation for the atom-dimer scattering amplitude.
Single (double) line indicates single atom (two-atom state).}
\label{fig:ineq}
\end{figure}

We now turn to the 3-body system. 
At leading order, the atom-dimer scattering amplitude is given by the 
integral equation shown in  Fig.~\ref{fig:ineq}. 
A solid line indicates a single atom and a double line indicates an 
interacting two-atom state (including rescattering corrections).
The integral equation contains contributions from both
the 2-body and the 3-body interaction terms.
The inhomogeneous term is given by the first two diagrams on the 
right-hand side: the one-atom exchange diagram
and the 3-body term. The integral equation simply sums
these diagrams to all orders.
After projecting onto S-waves, we obtain the equation
\begin{eqnarray}
{\cal T} (k, p; E) & = & {16 \over 3 a}\, M(k,p;E)
+ {4 \over \pi} \int_0^\Lambda 
{dq \, q^2 \, M(q,p;E)\over  -{1/a} + \sqrt{3q^2/4 -mE
-i \epsilon}}\, {\cal T} (k, q; E)\,,
\label{eq:BHvK}
\end{eqnarray}
for the off-shell atom-dimer scattering amplitude with the inhomogeneous
term 
\begin{eqnarray}
M(k,p;E)&=& {1 \over 2pk} \ln \left({p^2 + pk + k^2 -mE \over
p^2 - pk + k^2 - mE}\right) + {H(\Lambda) \over \Lambda^2}\,.
\end{eqnarray}
The first term is the S-wave projected one-atom exchange,
while the second term comes from the 3-body
interaction. The physical atom-dimer scattering amplitude 
$f_{AD}$ is given by the solution ${\cal T}$ evaluated at the on-shell 
point:~$f_{AD}(k) = {\cal T} (k, k; E)$ where
$E= 3k^2/(4m)-1/(ma^2)\,$.
The 3-body binding energies $B_3$ are given by those values of $E$
for which the homogeneous version of Eq.~(\ref{eq:BHvK}) has a
nontrivial solution.

Note that an ultraviolet cutoff $\Lambda$ has been introduced in
(\ref{eq:BHvK}). This cutoff is required to insure that Eq.~(\ref{eq:BHvK})
has a unique solution. All physical observables, however, must be invariant
under changes of the cutoff, which determines the behavior of $H$ as a 
function of $\Lambda$ \cite{BhvK99}:
\begin{eqnarray}
H (\Lambda) = {\cos [s_0 \ln (\Lambda/ \Lambda_*) + \arctan s_0]
\over \cos [s_0 \ln (\Lambda/ \Lambda_*) - \arctan s_0]}\,,
\label{H-Lambda}
\end{eqnarray}
where $s_0=1.00624$ is a transcendental number and $\Lambda_*$
is a 3-body parameter introduced by dimensional transmutation. 
This parameter cannot be predicted by the EFT and must be taken from 
experiment. It is evident that $H (\Lambda)$ is periodic and runs on
a limit cycle. When $\Lambda$ is changed by the preferred scaling factor
$\lambda=\exp(\pi/s_0)\approx 22.7$, $H$ returns to its original 
value. Note that the definition of the 3-body parameter is not 
unique. The parameter $\Lambda_*$ from Eq.~(\ref{H-Lambda})
arises naturally in the EFT description, while the parameter
$\kappa_*$ from Eq.~(\ref{kappa-star}) is defined via the 
3-body Efimov spectrum in the limit $a\to \pm \infty$. Both definitions 
are related by a constant factor: $\Lambda_* =2.62 \kappa_*$
\cite{Braaten:2004rn}.

In summary, two parameters are required in the 
3-body system at leading order in $l/a$:
the scattering length $a$ (or the dimer
binding energy $B_2$) and the 3-body parameter $\Lambda_*$ 
or $\kappa_*$ \cite{BhvK99}. 
The EFT reproduces the universal aspects of the 3-body system
first derived by Efimov \cite{Efi71} and is a very efficient
calculational tool to calculate those properties.

\section{APPLICATIONS TO COLD ATOMS}

\begin{figure}[htb]
\centerline{\includegraphics*[width=8.cm,angle=0]{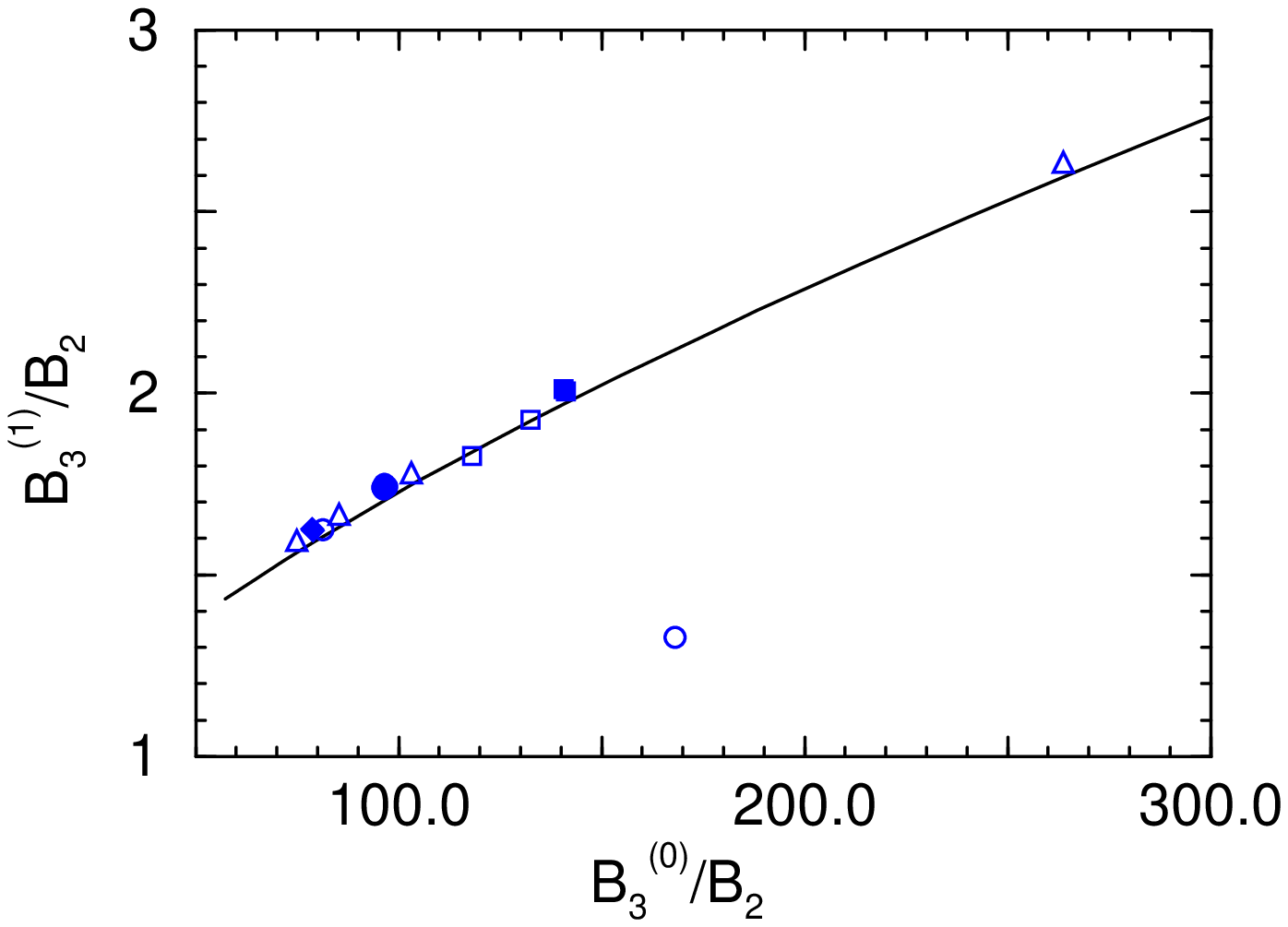}
\quad
\includegraphics*[width=7.5cm,angle=0]{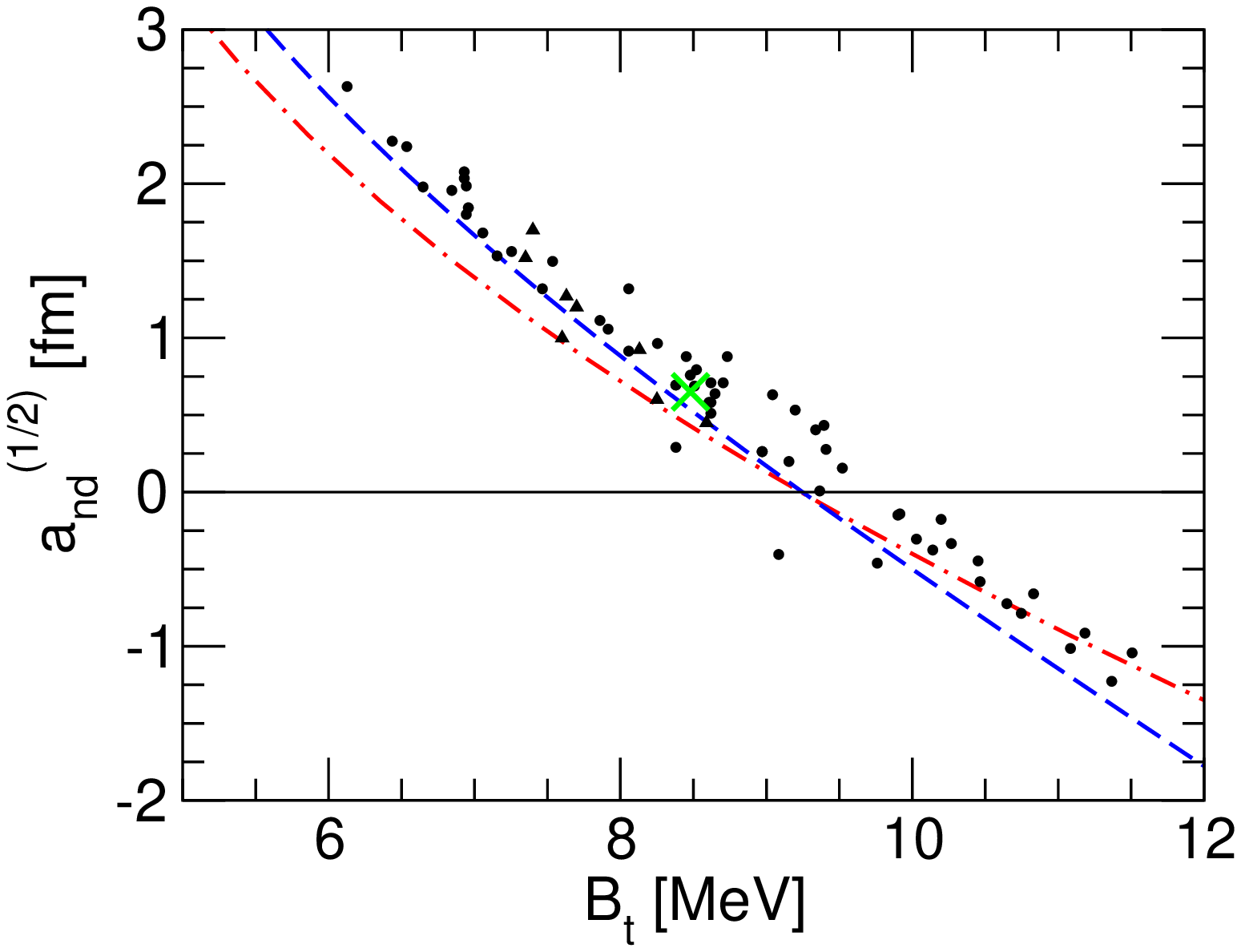}}
\caption{Left panel: The scaling function $B_3^{(1)}/B_2$ versus 
$B_3^{(0)}/B_2$ for $^4$He atoms \cite{he4univ}. 
Right panel: The scaling function
$a_{nd}^{(1/2)}$ versus $B_t$ from nuclear physics \cite{Bedaque:2002yg}. 
The dash-dotted and dashed lines exclude and include the leading 
correction in $l/a$, respectively. The data points are calculations
using realistic potentials.}
\label{fig:scaling}
\end{figure}

We now turn to some applications of this EFT to systems of cold atoms.
First we discuss universal scaling functions. Since only two
parameters enter at leading oder, different
3-body observables show correlations. These correlations
must appear in all 3-body systems with short-range
interactions and large scattering length.
In the left panel of Fig.~\ref{fig:scaling}, we display the
scaling function relating the trimer excited and ground state energies
$B_3^{(1)}$ and  $B_3^{(0)}$, respectively \cite{he4univ}.
The data points give various calculations using realistic $^4$He potentials 
while the solid line gives the universal prediction from EFT. 
Different points on this line correspond to different values of
$\kappa_*$. The small deviations of the 
potential calculations from the universal curve are mainly due to 
effective range corrections and can be calculated at next-to-leading
order in EFT \cite{PlaP06}. The calculation corresponding to the
data point far off the universal curve can easily be 
identified as problematic.
In the right panel of Fig.~\ref{fig:scaling}, we display
a similar correlation between the triton binding $B_t$ and the 
spin-doublet $S$-wave neutron-deuteron scattering length 
$a_{nd}^{(1/2)}$ from nuclear physics taken from \cite{Bedaque:2002yg}, 
which is known as the Phillips line.
The dash-dotted and dashed lines give the results excluding and
including the leading corrections in $l/a$, respectively.
It is evident that nuclear and atomic systems show the same universal 
scaling behavior. While these systems have very different 
structure at short distances, all that matters here is that $|a|\gg l$
but not the details of the 
mechanism leading to the large scattering length.

There are also observable features directly related to the 
occurence of a limit cycle in the 3-body system. 
The discrete scale invariance is manifest in the log-periodic 
dependence of 3-body observables on the scattering length $a$. This
dependence can be tested in experiments with cold atoms close to
a Feshbach resonance.\footnote{
See Ref.~\cite{Braaten:2003eu} for the conjecture of a limit cycle
in a deformed version of Quantum Chromodynamics.}
For this purpose we consider 3-body recombination, which is the process when
three atoms scatter to form a dimer and the third atom balances energy
and momentum. This is one of the main loss processes for trapped atoms
and condensates of atoms near a Feshbach resonance. The event rate $\nu$
per unit time and unit volume 
can be parametrized as $\nu=\alpha\rho^3$, where $\rho$ is the density
of the atoms and $\alpha$ is the recombination constant. 

Unfortunately, heavy alkali atoms, such as Rb and Cs, form
many deeply-bound diatomic molecules. Therefore, Efimov states
are resonances rather than sharp states, because they
can decay into a deep molecule and an energetic atom.
The presence of deep molecules also affects other 3-body observables,
but using unitarity their influence can be accounted for 
by one real parameter $\eta_*$ \cite{dimdeac}.
This parameter describes how much probability is lost by scattering
into the deep bound states. For $\eta_* =0$ the observables are
not modified, while for  $\eta_* =\infty$ all probability is lost
into the deep states. The consequences of the limit cycle become
totally washed out for $\eta_*\simge 1$.
The 3-body recombination coefficient $\alpha$ was first calculated
in Refs.~\cite{NM-99,EGB-99,3brec}. 
Analytical expressions for $\alpha$ as a function of $a$, $\kappa_*$,
and $\eta_*$ have been obtained in Refs.~\cite{dimdeac,Braaten:2004rn}.

In a recent experiment with cold $^{133}$Cs atoms,
the Innsbruck group has presented the first experimental evidence
for Efimov physics \cite{Grimm06}.
They used a Feshbach resonance to control the scattering length of
$^{133}$Cs atoms.
Since inelastic 2-body losses were energetically forbidden,
the dominant loss mechanisms were inelastic 3-body losses.
By varying the external magnetic field, they were able to change the
scattering length from $-2500 \ a_0$ through 0 to $+1600 \ a_0$,
where $a_0$ is the Bohr radius.
They observed a giant loss feature at a magnetic field that
corresponds to the scattering length $a = -850(20) \ a_0$.
They also measured the 3-body recombination rate for
positive values of $a$ reached by increasing the magnetic field
through a zero of the scattering length.
They observed a local minimum in the inelastic loss rate at a magnetic
field that corresponds to a scattering length of 210(10) $a_0$.
Their data could well described by the universal expressions from 
Refs.~\cite{dimdeac,Braaten:2004rn}, therefore giving the first 
indirect evidence for Efimov states in cold atoms \cite{Grimm06}.

\begin{figure}[htb]
\centerline{\includegraphics*[width=7cm,angle=0]{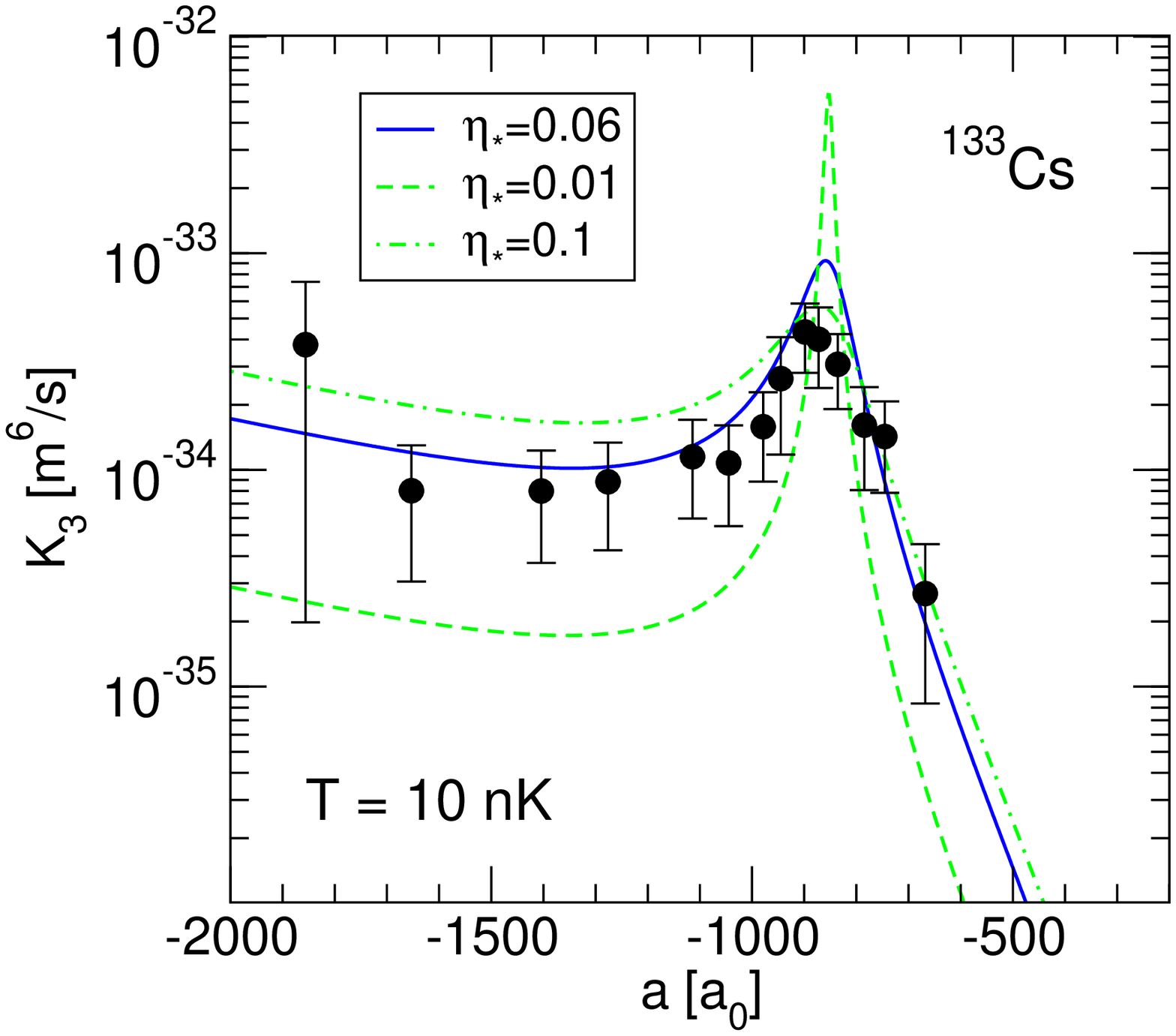}
\qquad
\includegraphics*[width=7cm,angle=0]{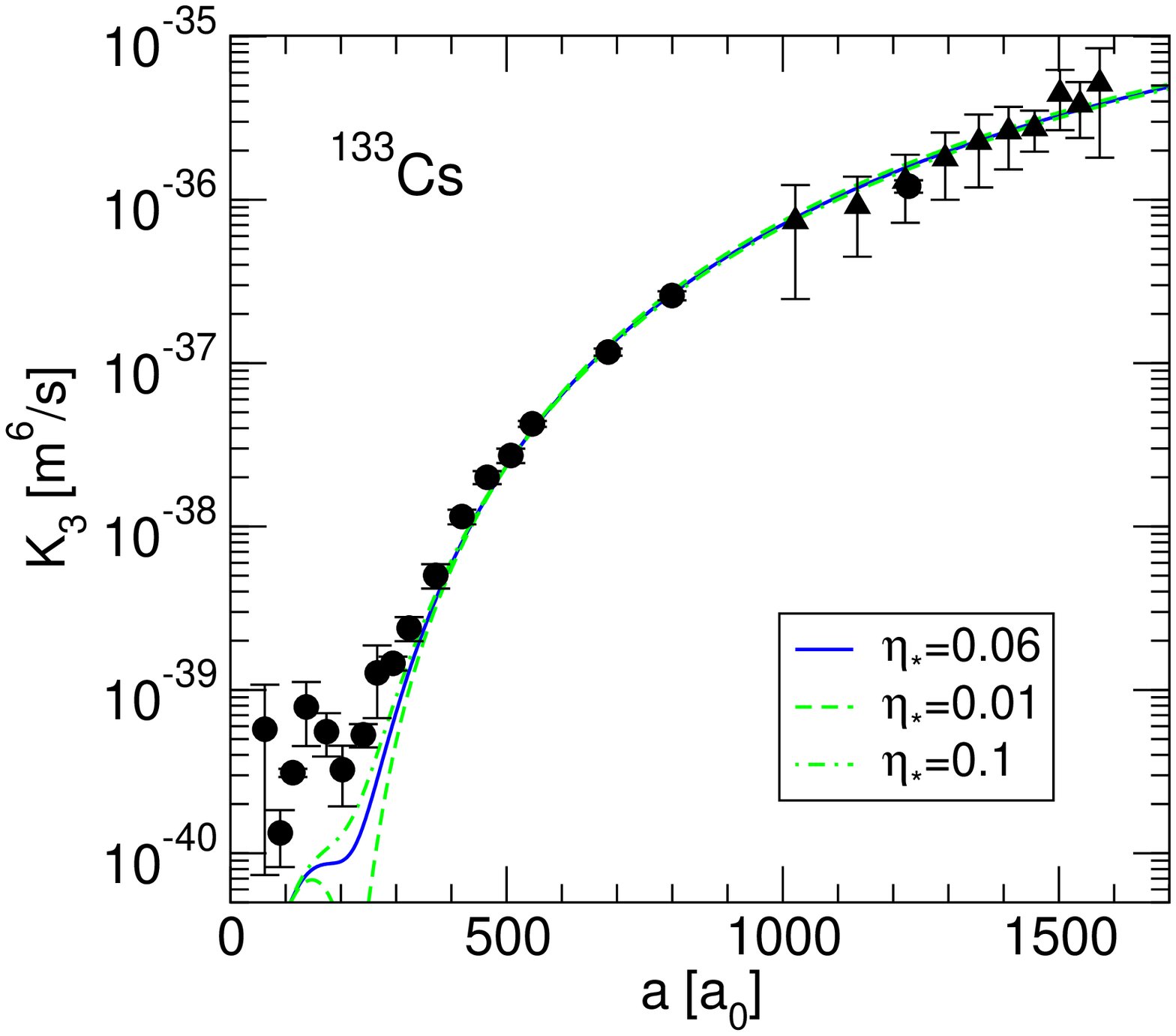}}
\caption{Left panel: 
The 3-body loss coefficient $K_3=3\alpha$ 
in $^{133}$Cs for $a<0$. The curves are for $\kappa_* = 0.945/a_0$ and
three different values of $\eta_*$. The data points are for $T=10$ nK 
\cite{Grimm06}.
Right panel: $K_3$ in $^{133}$Cs for $a>0$.  
The curves are for $\kappa_* = 0.707/a_0$ and three different values 
of $\eta_*$. The data points are for $T\approx 250...400$ nK \cite{Grimm06}.}
\label{fig:133Cs}
\end{figure}

In the left panel of Fig.~\ref{fig:133Cs}, we compare 
the universal prediction for negative scattering length 
at $T=0$ with the Innsbruck data 
for the loss coefficient $K_3 =3\alpha$ at $T=10$ nK \cite{Grimm06}.
For $\kappa_*=0.945/a_0$ and $\eta_*=0.06$, one obtains a good fit.
Both parameters are well determined by the resonance: $\kappa_*$
by the position of the Efimov resonance and $\eta_*$ by the height of the
peak.
The positive scattering length data 
from Ref.~\cite{Grimm06} are shown in the right panel 
of Fig.~\ref{fig:133Cs}. 
One complication here is that the lowest
temperature that was reached for positive $a$ was 250 nK.
To take account the thermal effects properly would require knowing
the 3-body recombination rate as a function of the collision energy.
We compare the universal expression for $T=0$ with
the data for $T\approx 250...400$ nK \cite{Grimm06}.
For $\kappa_*=0.707/a_0$ and  $\eta_*=0.06$, one
obtains a good fit to the data above $a\approx 400\,a_0$.
However,
the value of  $\eta_*$ is not well determined by the data and only the 
upper bound $\eta_* < 0.2$ can be given \cite{Grimm06}.

We note that the values of $\kappa_*$ and $\eta_*$ for positive and
negative scattering length need not be the same in the present case
since both regions are divided by a nonuniversal region with small
scattering length.
Moreover the universal expressions are rigorously only valid for
$|a| \gg l$ with $l=l_{vdW}\approx 200\; a_0$ for Cs atoms. 
Therefore the minimum around  $a=210(10) \;a_0$ is outside
the region of validity of the universal theory.

\section{SUMMARY \& OUTLOOK}

We have discussed the universal properties of few-body systems with large 
scattering length $a \gg l$. In the 3-body system these properties include
the Efimov effect and a discrete scaling symmetry leading to
log-periodic dependence of 3-body observables on 
the scattering length. These features can be understood as 
manifestations of a RG limit cycle in the 3-body system.
Moreover, we have presented an EFT that is designed to exploit the 
separation of scales in few-body systems with large scattering length.
In this EFT, the limit cycle is manifest through the RG 
behavior of the 3-body interaction required for proper 
renormalization. The EFT is very general and has applications
in few-body systems from atomic to nuclear and particle physics.
As an example, we have shown universal scaling functions for
$^4$He atoms and the Phillips line.
Furthermore, we have discussed the recent 3-body recombination data
for cold $^{133}$Cs atoms by the Innsbruck group that provided the first 
evidence of Efimov states in cold atoms \cite{Grimm06}.
Future challenges include the extension of the EFT to the four-body system
\cite{Platter:2004qn,Yama06}
and the proper calculation of finite temperature effects 
\cite{Jonsell06,YFT06,finiteT}.

This work was done in collaboration with Eric Braaten. It was supported 
by the DFG through SFB/TR 16 \lq\lq Subnuclear structure of matter''
and by the BMBF under contract number 06BN411.


\begin{thebibliography}{99}

\bibitem{Wilson:1970ag}
K.~G.~Wilson,
        Phys.\ Rev.\ D 3, 1818 (1971).

\bibitem{Efi71}
   V.N.\ Efimov, Sov.\ J.\ Nucl.\ Phys.\ 12 (1971) 589;
   29 (1979) 546.

\bibitem{Braaten:2004rn}
  E.~Braaten and H.-W.~Hammer,
  Phys.\ Rept.\  428 (2006) 259.

\bibitem{BhvK99}
 P.~F.~Bedaque, H.-W.~Hammer, and U.~van Kolck,
 Phys.\ Rev.\ Lett.\  82 (1999) 463;
  Nucl.\ Phys.\ A 646 (1999) 444.

\bibitem{he4univ}
E.~Braaten and H.-W.~Hammer,
Phys.\ Rev.\ A 67 (2003) 042706.

\bibitem{PlaP06}
L.~Platter and D.R.~Phillips,
arXiv:cond-mat/0604255.

\bibitem{Bedaque:2002yg}
  P.~F.~Bedaque, G.~Rupak, H.W.~Grie\ss hammer and H.-W.~Hammer,
  Nucl.\ Phys.\ A 714 (2003) 589.

\bibitem{Braaten:2003eu}
  E.~Braaten and H.-W.~Hammer,
  Phys.\ Rev.\ Lett.\  91 (2003) 102002.

\bibitem{dimdeac}
E.~Braaten and H.-W.~Hammer,
Phys.\ Rev.\ A 70 (2004) 042706.

\bibitem{NM-99}
E.~Nielsen and J.H.~Macek,
Phys.\ Rev.\ Lett.\ 83 (1999) 1566.

\bibitem{EGB-99}
B.D.~Esry, C.H.~Greene, and J.P.~Burke,
Phys.\ Rev.\ Lett.\ 83 (1999) 1751.

\bibitem{3brec}
P.~F.~Bedaque, E.~Braaten, and H.-W.~Hammer,
Phys.\ Rev.\ Lett.\ 85 (2000) 908;
E.~Braaten and H.-W.~Hammer,
Phys.\ Rev.\ Lett.\ 87 (2001) 160407.

\bibitem{Grimm06}
T. Kraemer, M.~Mark, P.~Waldburger, J.G.~Danzl, C.~Chin, B.~Engeser,
A.D.~Lange, K.~Pilch, A.~Jaakkola, H.-C.~N\"agerl, and R.~Grimm,
Nature 440 (2006) 315.

\bibitem{Platter:2004qn}
  L.~Platter, H.-W.~Hammer and U.-G.~Mei\ss ner,
  Phys.\ Rev.\ A 70 (2004) 052101.

\bibitem{Yama06}
M.T.~Yamashita, L.~Tomio, A.~Delfino, T.~Frederico,
Europhys.\ Lett.\ 75 (2006) 555.

\bibitem{Jonsell06}
S.~Jonsell,
Europhys.\ Lett.\ 76 (2006) 8.

\bibitem{YFT06}
M.T.~Yamashita, T.~Frederico, and Lauro Tomio,
arXiv:cond-mat/0608542.

\bibitem{finiteT}
E.~Braaten and H.-W.~Hammer,
arXiv:cond-mat/0610116.

\end{thebibliography}
\end{document}